\documentclass[a4paper,11pt]{article}
\usepackage[utf8x]{inputenc}
\usepackage{amsmath,amssymb,amsfonts}
\usepackage[english]{babel}
\usepackage{multirow}
\usepackage{bigdelim}
\usepackage{xspace}
\usepackage[small, bf, hang]{caption}
\usepackage[numbers]{natbib}
\usepackage[hypertex]{hyperref}
\usepackage{hypernat}

\newcommand{\rep}[1]{\ensuremath{\mathbf{#1}}}
\newcommand{\crep}[1]{\ensuremath{\mathbf{\overline{#1}}}}
\newcommand{\hc}{\ensuremath{\text{h.c.}}}
\newcommand{\im}{\ensuremath{\text{i}}}
\DeclareMathOperator{\tr}{tr}
\DeclareMathOperator{\diag}{diag}

\newcommand{\dorsner}{Dor\v{s}ner\xspace}
\newcommand{\perez}{P\'{e}rez\xspace}

\begin{document}
\begin{flushright}
SI-HEP-2012-08
\end{flushright}
\bigskip
\begin{center}
{\Large\bfseries%
Higgs Mass Spectrum in non-SUSY SU(5)}\\[2ex]
Karsten Schnitter\\[1ex]
{\itshape Theoretische Elemtarteilchenphysik, Naturwissenschaftlich Technische Fakult\"at,\\
Universit\"at Siegen, 57068 Siegen, Germany}
\end{center}
\bigskip
\begin{abstract}
Non-supersymmetric grand unified theories based on SU(5) have had a revival during the past years.
This is mainly due to their ability to connect neutrino masses with unification and proton decay.
In that way they provide a framework for testable models at coming experiments. 
We study the mass spectrum of such models and determine naturally light multiplets within the LHC reach.
\end{abstract}
\bigskip

\section{Introduction}
Grand unified theories belong to the best studied extensions of the standard model (SM).
They offer a powerful framework in which different problems of the SM can be addressed.
Especially appealing are the gauge coupling unification and the relations between different Yukawa couplings.
Both effects are direct consequences of the embedding of the SM gauge group $SU(3)_C\times SU(2)_L \times U(1)_Y$ into a larger GUT group for which SU(5) is the smallest possible choice \cite{Georgi:1974sy}.
The SU(5) also appears as a subgroup of other popular GUT groups such as SO(10) or $\text{E}_6$.

A further advantage of GUTs is the possibility of the combination with supersymmetry (SUSY).
For SU(5) with minimal field content SUSY helps to evade the stringent bounds set by proton decay that excludes the non-SUSY model.
However, there is no a-priori reason for the introduction of SUSY if we can formulate other models that accommodate the experimental results.

The simplest extension of the minimal Georgi-Glashow SU(5) is the addition of a scalar 15-plet.
The resulting model has a number of interesting phenomenological consequences, e.g. a Type-II-see-saw neutrino mass generation or a possibly light leptoquark \cite{Dorsner:2005fq, Dorsner:2005ii, Dorsner:2006dj, Dorsner:2006hw}.
We investigate the symmetry breaking within this model and determine the viable mass ranges of all scalar particles in this paper.
Since this model incorporates two very separate mass scales splitting SU(5) multiplets into both heavy and light fields, fine-tuning is unavoidable.
We will show that a light leptoquark requires additional fine-tuning to the always present fine-tuning needed to obtain a light SM Higgs field.

In \autoref{sec:model} we introduce the model and the notation.
Furthermore we define two special cases for the low energy effective theory.
\hyperlink{sec:spectrum}{Section~\ref{sec:spectrum}} covers the mass determination for the heavy fields and explains the fine-tunings leading to small masses for all other fields.
The easier of the two special cases from \autoref{sec:model} is discussed in \autoref{sec:sim-limit} and the decoupling of the non-SM like fields from the SM is demonstrated. 
\hyperlink{sec:cust-limit}{Section~\ref{sec:cust-limit}} continues this discussion for the less restrictive second special case.
This leads to possibly light scalar particles for which lower mass bounds are determined in a numerical simulation.
The final conclusion can be found in \autoref{sec:conclusion}.
\autoref{app:potential} contains the explicit formulas for all potentials and mass matrices.
\pagebreak

\section{The Model} \label{sec:model}
We consider a non-SUSY SU(5) model proposed by \dorsner and \perez \cite{Dorsner:2005fq}. 
It consists of the minimal model from Georgi and Glashow \cite{Georgi:1974sy} but is extended by an additional scalar 15-plet. 
The full scalar field content of this model and the decomposition of the SU(5) multiplets under the SM gauge subgroup can be found in \autoref{tab:higgses}.
We use these multiplets to construct the most general renormalizable potential $\mathcal{V}$ in \autoref{app:potential}.
\begin{table}[h]
\centering
  \begin{tabular}{lll}
    \hline
      \multicolumn{2}{c}{SU(5) multiplet} & Notation\\
    \hline\\[-1.5ex]
      \ldelim\lbrace{2}{5ex}[\rep{\phantom{9}5}\ ] & $(\rep{3},\rep{1})_{-1/3}$ & \ $\Phi_3$\\
      & $(\rep{1},\rep{2})_{1/2}$& \ $\Phi_2$\\[0.5ex]
    \hline\\[-1.5ex]
      \ldelim\lbrace{3}{5ex}[\rep{15}\ ] & $(\rep{6},\rep{1})_{-2/3}$ & \ $\Upsilon_6$\\
      & $(\rep{3},\rep{2})_{1/6}$ & \ $\Upsilon_{LQ}$\\
      & $(\rep{1},\rep{3})_{1}$ & \ $\Upsilon_{3}$\\[0.5ex]
    \hline\\[-1.5ex] 
      \ldelim\lbrace{5}{5ex}[\rep{24}\ ] & $(\rep{8},\rep{1})_1$ & \ $\Sigma_8$\\
      & $(\rep{3},\rep{2})_{-5/6}$ & \ $\Sigma_{NGB}$ \\ 
      & $(\crep{3},\rep{2})_{5/6}$ & \ $\overline{\Sigma}_{NGB}$\\
      & $(\rep{1},\rep{3})_0$ & \ $\Sigma_3$ \\
      & $(\rep{1},\rep{1})_0$ & \ $\Sigma_1$\\[0.5ex]
    \hline
  \end{tabular}
  \caption{SM decompositon of SU(5) multiplets, SM fields are noted by their gauge multiplets
$(C,L)_Y$ of $G_\text{SM}=SU(3)_C\times SU(2)_L \times U(1)_Y$}
  \label{tab:higgses}
\end{table}

It is evident from \autoref{tab:higgses}, that the SU(5) breaking can only be archieved by a non-vanishing vacuum expectation value (VEV) $\langle\Sigma_1\rangle = v_S \ne 0$. 
However for the electroweak symmetry breaking (EWSB) VEVs of the neutral components of the fields $\langle\Phi_2^0\rangle=v_D$, $\langle\Sigma_3^0\rangle=v_T$ and $\langle\Upsilon_3^0\rangle=v_\Upsilon$ can contribute.
For details of the potential and the definition of the VEVs see \autoref{app:potential}.

Using these definitions we can determine the gauge boson masses. 
Gluons and Photon stay exactly massless, the mass of the other bosons are given in \eqref{eqn:gauge-masses}.
\begin{subequations} \label{eqn:gauge-masses}
\begin{align} 
\label{eqn:gauge-masses-gut}
  M_X^2 &= \frac{g^2}{4} (5v_S-v_T)^2 &
  M_Y^2 &= \frac{g^2}{4} \bigl((5v_S-v_T)^2+v_D^2+4v_\Upsilon^2\bigr)\\
\label{eqn:gauge-masses-ewsb}
  M_Z^2 &= \frac{g^2}{4}\,\frac{8}{5} \bigl(v_D^2+8v_\Upsilon^2\bigr) &
  M_W^2 &= \frac{g^2}{4} \bigl(v_D^2+4(v_T^2+v_\Upsilon^2)\bigr)
\end{align}
\end{subequations}
Since the X- and Y-bosons mediate proton decay, there is a huge hierarchy between their masses and
the W- and Z-mass.
This hierarchy is carried by the VEVs.
We use a supression factor $\epsilon$ to indicate this hierarchy: $v_D,v_T,v_\Upsilon\sim\epsilon\,
v_S$.

From the masses of the W- and Z-boson given in \eqref{eqn:gauge-masses-ewsb} we can deduce the
tree-level rho-parameter in this model.
It needs to be compared to the experimental fit after separating SM loop contributions.
In \eqref{eqn:rho-parameter} we give the expression and the $2\sigma$ fit value from \cite{Nakamura:2010zzi}.
\begin{equation} \label{eqn:rho-parameter}
  \rho_\text{low}^\text{tree} = 
    \frac{v_D^2+4(v_T^2+v_\Upsilon^2)}{v_D^2+8v_\Upsilon^2}
    = 1 + 4 \frac{v_T^2-v_\Upsilon^2}{v_D^2+8v_\Upsilon^2}
    \approx 1.0008^{+0.0029}_{-0.0011}
\end{equation}
To meet this experimental constraint we investigate two possible limits:
\begin{enumerate}
  \item the simultaneous limit $v_T \to 0$ and $v_\Upsilon \to 0$ (c.f. \autoref{sec:sim-limit})
  \item the custodial limit $v_T^2-v_\Upsilon^2 \to 0$ (c.f. \autoref{sec:cust-limit})
\end{enumerate}

\section{Heavy Mass Spectrum} \label{sec:spectrum}
We now turn to the task of determining the mass spectrum of the heavy particles. 
As a first step we consider the minimisation conditions. 
These are only non-vanishing for field directions along the VEVs. 
Evaluating the derivative of the potential w.r.t. $\Sigma_1$ we obtain \eqref{eqn:vs-fixing}.
This condition can be used to fix $v_S$.
\begin{align} \label{eqn:vs-fixing}
  0 &= \frac{\partial\mathcal{V}}{\partial\Sigma_1}\Biggr\rvert_{\Sigma_1=v_S}\hspace{-2em} &
  &\leadsto &
  0 &= \frac{v_S}{4} \bigl(15a_1v_S^2 + 7b_1v_S^2 - 3c_1v_S - 4 m_1^2\bigr) 
      + \mathcal{O}(\epsilon^2v_S^3)
\end{align}
Similarly we obtain \eqref{eqn:ewsb-min-cond} form the minimisation conditions along the EWSB VEVs using only the dimensionful parameters defined in \eqref{eqn:ewsb-min-param}. 
\begin{subequations} \label{eqn:ewsb-min-cond}
\begin{align}
  &\langle\Phi_2^0\rangle: &
    0 &= \frac{v_D}{4} m_D^2 
      - \frac{v_Dv_T}{2} \mu_1 + \bigl(\frac{v_Dv_\Upsilon}{2}\mu_3 + \hc\bigr)
      + \mathcal{O}(\epsilon^3v_S^3)\\
  &\langle\Sigma_3^0\rangle: &
    0 &= \frac{v_T}{4} m_T^2 
      - v_D^2 \mu_1 -\frac{v_\Upsilon^2}{2}\mu_2
      + \mathcal{O}(\epsilon^3v_S^3)\\
  &\langle\Upsilon_3^0\rangle: &
    0 &= \frac{v_\Upsilon}{4} m_\Upsilon^2 
      + v_D^2 \mu_3 - \frac{v_Tv_\Upsilon}{2}\mu_2
      + \mathcal{O}(\epsilon^3v_S^3)
\end{align}
\end{subequations}
\begin{subequations} \label{eqn:ewsb-min-param}
\begin{align}
  m_D^2 &= 30 a_2 v_S^2 + 9 b_2 v_S^2 - 6 c_2 v_S -4 m_2^2\\
  m_T^2 &= 15 a_1 v_S^2 + 27 b_1 v_S^2 -18 c_1 v_S - 4 m_1^2\\
  m_\Upsilon^2 &= 30 a_5 v_S^2 + 9 (b_5+b_6) v_S^2 - 6 c_5 v_S - 4 m_3^2
\end{align}
\vspace{-4ex}
\begin{align}
  \mu_1 &= c_2 - 3 b_2 v_S &
  \mu_2 &= c_5-3(b_5+b_6)v_S &
  \mu_3 &= c_4-3b_7v_S
\end{align}
\end{subequations}

The natural scale of the parameters in \eqref{eqn:ewsb-min-param} is the SU(5) scale $v_S$. 
However, to achieve EWSB within this model the conditions \eqref{eqn:ewsb-min-cond} must be met. 
If that is done without further restrictions on the dimensionful parameters in \eqref{eqn:ewsb-min-param}, we will call this case \emph{required fine-tuning}.
This name reflects the need for fine-tuning because of the large hierarchy $\epsilon$ between SU(5) and EWSB scale.

Another interesting case can be obtained by restricting all dimensionful parameters to be of order of the EWSB scale $\epsilon\,v_S$.
This requirement is sufficient in the sense that the minimisation conditions \eqref{eqn:ewsb-min-cond} still pose restrictions but no longer contain any fine-tuning of order $\epsilon$.
Due to this fact we will call this case \emph{sufficient fine-tuning} for further reference.

Note that one could try to meet the minimisation conditions \eqref{eqn:ewsb-min-cond} by suppressing all appearing parameters by $\epsilon$.
Since the whole parameter space of the model is contained in these conditions, this does not lead to the desired hierarchy.

Continuing the determination of the scalar masses we consider the second derivatives of the SU(5) potential $\mathcal{V}$.
Using the minimisation conditions \eqref{eqn:ewsb-min-cond} we can prove, that the SU(5)-Nambu-Goldstone bosons $\Sigma_{NGB}$ remain exactly massless. 
After EWSB $\Sigma_{NGB}$ and $\Phi_3$ mix, but the mixing angle is strongly suppressed by $\epsilon$. 
The EWSB-NGBs can be obtained in the mixing of $\Phi_2$, $\Sigma_3$ and $\Upsilon_3$. 
Their mixing angles are given by the ratios of their respective VEVs $v_D$, $v_T$ and $v_\Upsilon$. 
Hence, this mixing may be suppressed as well as it may be maximal depending on those ratios.
See sections~\ref{sec:sim-limit} and \ref{sec:cust-limit} for further discussion.

Following the same road we can determine the complete mass spectrum of the model. 
In \eqref{eqn:heavy-masses} the masses of the SM singlet and all coloured fields are given.
\begin{subequations} \label{eqn:heavy-masses}
\begin{align}
\label{eqn:heavy-masses-1}
  M_{\Sigma_1}^2 &= 
    \frac{5}{2} (3 a_1+b_1) v_S^2 
      + \mathcal{O}(\epsilon\,v_S^2)\\
  M_{\Sigma_8}^2 &= 
    \frac{25}{4} b_1 v_S^2 
    -\mu_1 \frac{v_D^2}{v_T} 
    -\frac{\mu_2}{2} \frac{v_\Upsilon^2}{v_T} 
      + \mathcal{O}(\epsilon^2v_S^2)\\
  M_{\Phi_3}^2 &= 
    \frac{25}{4} b_2 v_S^2 
    +\frac{5}{2} \mu_1 v_S 
      + \mathcal{O}(\epsilon\,v_S^2)\\
\label{eqn:heavy-masses-4}
  M_{\Upsilon_6}^2 &= 
    \frac{25}{4} (b_5+b_6) v_S^2 
    +\frac{5}{2} \mu_2 v_S 
      + \mathcal{O}(\epsilon\,v_S^2)\\
\label{eqn:heavy-masses-5}
  M_{\Upsilon_{LQ}}^2  &= 
    \frac{25}{8} b_5 v_S^2 
    +\frac{5}{4} \mu_2 v_S
      + \mathcal{O}(\epsilon\,v_S^2)
\end{align}
\end{subequations}
These relations are rewritten using the minimisation conditions \eqref{eqn:ewsb-min-cond} to contain only dimensionful parameters as defined in \eqref{eqn:ewsb-min-param}.
Assuming the case of sufficient fine-tuning the generic scale for those masses clearly is the SU(5) scale $v_S$.
This remains true if we relax our claim to the case of required fine-tuning, but is no longer as easy to infer from \eqref{eqn:heavy-masses}. 

From the physics point of view $\Sigma_1$ is the Higgs boson of the SU(5) breaking.
Hence its mass is connected to the breaking scale $v_S$ and correctly described by \eqref{eqn:heavy-masses-1}.
The impact of the remaining fields in \eqref{eqn:heavy-masses} on unification and proton decay was studied in great detail in \cite{Dorsner:2005fq, Dorsner:2005ii, Dorsner:2006dj, Dorsner:2006hw}.
The main result was that $\Sigma_8$, $\Phi_3$ and $\Upsilon_6$ need to obtain heavy masses of the order of the SU(5) scale $v_S$ while for the leptoquark $\Upsilon_{LQ}$ a rather light mass within the LHC reach is preferred.
This implies, that the mass spectrum of the 15-plet $\Upsilon$ needs to be split into a heavy part containing the sixtet $\Upsilon_6$ and a light part containing $\Upsilon_{LQ}$ similarly to the doublet-triplet-splitting (DTS) of the quintet $\Phi$ into $\Phi_3$ and $\Phi_2$.
Combining \eqref{eqn:heavy-masses-4} and \eqref{eqn:heavy-masses-5} to \eqref{eqn:15-splitting} we see, that the splitting of the 15-plet into $\Upsilon_6$ and $\Upsilon_{LQ}$ requires $b_6\sim\mathcal{O}(1)$.
This disables the possibility to choose $b_5$, $b_6$ and $c_5$ small by some symmetry in order to obtain a small leptoquark mass.
Hence a light leptoquark requires an additional fine-tuning within this model independent whether DTS is achieved by the required or sufficient fine-tuning mentioned earlier.
\begin{equation} \label{eqn:15-splitting}
  M_{\Upsilon_6}^2 = 2\,M_{\Upsilon_{LQ}}^2 + \frac{25}{4} b_6v_S^2 + \mathcal{O}(\epsilon\,v_S^2)
\end{equation}

The only fields still unattended are the weak isospin doublet $\Phi_2$ and the triplets $\Sigma_3$ and $\Upsilon_3$. 
They can be described by an effective two-Higgs-triplet model (2HTM) with the potential given in 
\eqref{eqn:potential-2htm}.
In this model all field components given in \eqref{eqn:heavy-masses} are integrated out and only the dimensionful parameters defined in \eqref{eqn:ewsb-min-param} are used.
Considering the case of sufficient fine-tuning previously is therefore equivalent to EWSB-scale 2HTM.
Note that the minimisation conditions of the 2HTM lead to same constraints \eqref{eqn:ewsb-min-cond} for the EWSB VEVs as the full SU(5) theory.

There have been extensive studies of models with two Higgs triplets \cite{Georgi:1985nv, Chanowitz:1985ug, Gunion:1989ci, Gunion:1990dt} based on a custodial SU(2) symmetry that ensures $\rho = 1$. 
In our model we will impose this symmetry only for the VEVs but consider general potential parameters.
We will continue the determination of the mass eigenstates of this theory in the following sections.
The main difficulty of this task is the mixing of states of same electric charge leading to the mass matrices given in \autoref{app:potential}.
The only exception is the mass of the doubly charged components $\Upsilon_3^{\pm\pm}$ which cannot mix with any other field:
\begin{subequations}\label{eqn:double-charged-mass}
\begin{align}
  M_{\Upsilon_3^{\pm\pm}}^2 
  &= \label{eqn:double-charged-mass-massterm}
    \tfrac{1}{4} m_\Upsilon^2
    + \tfrac{1}{2} v_T \Bigl(\mu_2 + \tfrac{v_T}{2}\bigl(2 a_5+b_5+b_6\bigr)\Bigr)
    + 2 v_D^2 a_4
    + \tfrac{1}{2} v_\Upsilon^2 a_3\\ 
  &= \label{eqn:double-charged-mass-seesaw}
    \frac{v_D^2}{v_\Upsilon} \bigl(b_7 v_T - \mu_3\bigr)
    + v_T \mu_2 
    - 2 v_D^2 b_4
    - v_\Upsilon^2 b_3
\text{.}
\end{align}
\end{subequations}
Note that this mass depends on the SU(5) scale only through the dimensionful parameters \eqref{eqn:ewsb-min-param} that determine the validity of the effective 2HTM approach.
These two expressions in \eqref{eqn:double-charged-mass} differ only by application of \eqref{eqn:ewsb-min-cond}, showing in \eqref{eqn:double-charged-mass-seesaw} a see-saw like dependence of the mass on the VEV ratio.
This demonstrates that the non-SM like field $\Upsilon_3^{\pm\pm}$ decouples in the limit $v_\Upsilon\to0$.
Given the case of sufficient fine-tuning this would not show up in \eqref{eqn:double-charged-mass-massterm} without the minimisation conditions properly taken into account.
We will see that this behavior is a prototype for all non-standard model fields of the 2HTM in the next sections.

Furthermore we can infer from leading terms in hierarchy $\epsilon$ of \eqref{eqn:double-charged-mass-massterm}:
\begin{equation}
  M_{\Upsilon_3^{\pm\pm}} \sim v_S \sqrt{\epsilon}\ \sqrt{\frac{v_D}{v_\Upsilon}+\frac{v_T}{v_D}}
\end{equation}
Hence for $\epsilon \ll v_\Upsilon/v_D \ll 1$ the doubly charged scalar $\Upsilon_3^{\pm\pm}$ acquires mass firmly below the SU(5) scale.
With the same argument as was done for $\Upsilon_6$ and $\Upsilon_{LQ}$ before \eqref{eqn:15-splitting} this leads to a splitting of the 15-plet $\Upsilon$.
But here this splitting is the consequence of the fine-tuning used to solve the DTS problem.
In that case a light weak isospin triplet $\Upsilon_3$ is more natural than a light leptoquark $\Upsilon_{LQ}$, since it does not require additional fine-tuning.

\section{Light Masses in the Simultaneous Limit} \label{sec:sim-limit}
We now discuss the first solution to the experimental constraint of the tree-level rho-parameter $v_T,v_\Upsilon\to0$.
To parametrize the simultaneous limit we introduce a new suppression parameter $\delta$ between the doublet and the triplet VEVs: $v_T,v_\Upsilon \sim \delta\,v_D$.
Given the experimental bounds in \eqref{eqn:rho-parameter} we get the upper bound $\delta < 10^{-1}$.

In leading order in $\delta$ (and $\epsilon$) we observe in \eqref{eqn:SL-cust-decomp} that there is a one-to-one correspondence of the mass eigenstates to the fields of our effective model since all mixings are suppressed by $\delta$.
Note that the right column also applies to the Georgi-Glashow model without the additional 15-plet contribution of $\Upsilon_3$%
\footnote{In the Georgi-Glashow model $f=1$ can be obtained}.
\begin{gather} \label{eqn:SL-cust-decomp}
\begin{aligned}
  H_5^{\pm\pm} &= \Upsilon_3^{\pm\pm} \\
  H_5^\pm &= \Upsilon_3^\pm - \tfrac{v_\Upsilon}{v_D} \Phi_2^\pm \\
  H_5^0 &= \text{Re}\,\Upsilon_3^0 - \tfrac{v_\Upsilon}{v_D}\sqrt{2}\,\text{Re}\,\Phi_2^0 \\
  A^0 &= \text{Im}\,\Upsilon_3^0 - \tfrac{v_\Upsilon}{v_D}\sqrt{2}\,\text{Im}\,\Phi_2^0 \\
\end{aligned}\qquad
\begin{aligned}
  H_3^\pm &= \mp\im f \Sigma_3^0 + \tfrac{v_T}{v_D} f \Phi_2^\pm\\
  H_3^0 &= \Sigma_3^0 - \tfrac{v_T}{v_D}\,\text{Re}\,\Phi_2^0\\
  h^0 &= \text{Re}\,\Phi_2^0 + \tfrac{v_T}{v_D} \Sigma_3^0 + \tfrac{v_\Upsilon}{v_D}\sqrt{2}\,\text{Re}\,\Upsilon_3^0
\end{aligned}\\
\text{with}\ 
 f = \frac{2v_T\mu_3 + v_\Upsilon (2\mu_1 + 2v_Tb_4 + 2v_\Upsilon b_7)}{2v_T\mu_3 + v_\Upsilon (\mu_1 + 2v_Tb_4 + v_\Upsilon b_7)}\text{.}\nonumber
\end{gather}

The notation of the mass eigenstates reflect the multiplets of the custodial SU(2), e.g.\ the quintuplet $H_5^{++},H_5^+,H_5^0,H_5^-,H_5^{--}$.
Their masses are given in \eqref{eqn:SL-masses} expanded up to order $\delta^2$.
We see that the masses of quintuplet and triplet are quasi-degenerate with splittings suppressed by $\delta$.
\begin{subequations} \label{eqn:SL-masses}
\begin{align} 
  M_{H_5^{\pm\pm}}^2 &= 
    \frac{v_D^2}{v_\Upsilon} \bigl(b_7 v_T - \mu_3\bigr)
    + v_T \mu_2 
    - 2 v_D^2 b_4
    + \mathcal{O}(\delta^2v_D^2)
  \\
  M_{H_5^\pm}^2 &= 
    \frac{v_D^2}{v_\Upsilon} \bigl(b_7 v_T - \mu_3\bigr)
    + \frac{1}{2}v_T \mu_2
    - v_\Upsilon \mu_3
    - v_D^2 b_4
    + \mathcal{O}(\delta^2v_D^2)
  \\
  M_{H_5^0}^2 &=     
    \frac{v_D^2}{v_\Upsilon} \bigl(b_7 v_T - \mu_3\bigr)
    - 2v_\Upsilon \mu_3
    + \mathcal{O}(\epsilon^3v_S^3 + \delta^2v_D^2)
  \\
  M_{A^0}^2 &=     
    \frac{v_D^2+v_\Upsilon^2}{v_\Upsilon} (b_7 v_T - \mu_3)
  \\
  M_{H_3^\pm}^2 &= 
    \frac{v_D^2}{v_T} \bigl(\mu_1 + 2v_\Upsilon b_7\bigr)
    + v_T \mu_1 
    + \frac{1}{2} \frac{v_\Upsilon^2}{v_T} \mu_2
    + \mathcal{O}(\delta^2v_D^2)
  \\
  M_{H_3^0}^2 &= 
    \frac{v_D^2}{v_T} \bigl(\mu_1 + 2v_\Upsilon b_7\bigr)
    + v_T \mu_1 
    + \frac{v_\Upsilon^2}{v_T} \mu_2
    + \mathcal{O}(\epsilon^3v_S^3 + \delta^2v_D^2)
  \\
  M_{h^0}^2 &= 
    2 v_D^2 l
    - v_T \mu_1 
    + 2v_\Upsilon \mu_3
    + \mathcal{O}(\epsilon^3v_S^3 + \delta^2v_D^2)
  \nonumber\\
  &= -\tfrac{1}{2} m_D^2
    + \mathcal{O}(\epsilon^3v_S^3 + \delta^2v_D^2)
\end{align}
\end{subequations}
All masses except for the last one are proportional to $1/\delta$.
This last field $h^0$ corresponds to the SM Higgs and has the known mass relation whithin this limit.
For all other fields their see-saw like mass terms clearly lead to a decoupling of those fields in the limit where $\delta\to0$.
Even in the case of sufficient fine-tuning where $\mu_i=\mathcal{O}(\epsilon\,v_S)$ their masses are of order $\sim v_D/\sqrt{\delta}$ which with the current data amounts to roughly 2-3 TeV.
This can be considered a very soft lower limit with smaller masses possible if either of $\mu_1$ or $\mu_3$ was even smaller.
However, if one finds these particles one could use their mass splittings to determine $\mu_2$ and $\mu_3$ from \eqref{eqn:SL-masses}.

The NGBs are independent of any limit since they are exactly massless.
Indeed they can easily be obtained as zero modes of the corresponding mass matrices \eqref{eqn:mass-matrices}.
Their field decomposition is given in \eqref{eqn:SL-NGB-decomp}.
\begin{align} \label{eqn:SL-NGB-decomp}
  G^\pm &= v_D \Phi_2^\pm \mp \im v_T \Sigma_3^\pm + v_\Upsilon \Upsilon_3^\pm&
  G^0 &= \tfrac{v_D}{\sqrt{2}}\,\text{Im}\,\Phi_2^0 + v_\Upsilon\ \text{Im}\,\Upsilon_3^0
\end{align}
Note that in the simultaneous limit they correspond to the SM NGBs up to corrections of order $\delta$.

\section{Light Masses in the Custodial Limit} \label{sec:cust-limit}
We now turn to the much more interesting case in which the rho-parameter constraint is obeyed by choosing the difference between the triplet VEVs $v_T$ and $v_\Upsilon$ to be small.
The phenomenological implications for the LHC of the SM-like CP-even Higgs boson $h$ in this model have been studied in \cite{Logan:2010en}.
For the case of exact custodial symmetry in the VEVs, i.e.\ $v_T=v_\Upsilon\equiv v_C$ we can easily determine all mass eigenstates except for the neutral CP-even fields.
Their decomposition into the 2HTM multiplets are given in \eqref{eqn:CL-cust-decomp}.
Their respective masses are calculated in \eqref{eqn:CL-cust-masses}.
\begin{subequations} \label{eqn:CL-cust-decomp}
\begin{align}
  H^{\pm\pm} &= \Upsilon_3^{\pm\pm}\\
  H_1^\pm &= 
    2\frac{v_C^2}{v_D} \Bigl(\tilde{\mu}_1 v_C - (\tilde{\mu}_2+\tilde{\mu}_3)v_C - 2\tilde{m}^2\Bigr)\Phi_2^\pm
    \nonumber\\&\quad
    \pm\im\,\Bigl(\!(3\tilde{\mu}_1-\tilde{\mu}_2+\tilde{\mu}_3)v_C^2 + (\tilde{\mu}_1+\tilde{\mu}_3)v_D^2 - 2\tilde{m}^2v_C\Bigr)\Sigma_3^\pm
    \nonumber\\&\quad
    +\ \Bigl(\!(\tilde{\mu}_1+\tilde{\mu}_2+3\tilde{\mu}_3)v_C^2 + (\tilde{\mu}_1+\tilde{\mu}_3)v_D^2 + 2\tilde{m}^2v_C\Bigr)\Upsilon_3^\pm
  \\
  H_2^\pm &= 
    2\frac{v_C^2}{v_D} \Bigl(\tilde{\mu}_1 v_C - (\tilde{\mu}_2+\tilde{\mu}_3)v_C + 2\tilde{m}^2\Bigr)\Phi_2^\pm
    \nonumber\\&\quad
    \pm\im\,\Bigl(\!(3\tilde{\mu}_1-\tilde{\mu}_2+\tilde{\mu}_3)v_C^2 + (\tilde{\mu}_1+\tilde{\mu}_3)v_D^2 + 2\tilde{m}^2v_C\Bigr)\Sigma_3^\pm
    \nonumber\\&\quad
    +\ \Bigl(\!(\tilde{\mu}_1+\tilde{\mu}_2+3\tilde{\mu}_3)v_C^2 + (\tilde{\mu}_1+\tilde{\mu}_3)v_D^2 - 2\tilde{m}^2v_C\Bigr)\Upsilon_3^\pm
  \\
  A^0 &= \text{Im}\,\Upsilon_3^0 - \tfrac{v_\Upsilon}{v_D}\sqrt{2}\,\text{Im}\,\Phi_2^0
\end{align}
\end{subequations}
\begin{subequations} \label{eqn:CL-cust-masses}
\begin{align}
  M_{H^{\pm\pm}}^2 &=
    \frac{v_D^2}{v_C} \bigl(b_7 v_C - \mu_3\bigr)
    + v_C \mu_2
    - v_C^2 (b_3 + 2 b_4)\\
  M_{H_1^\pm}^2 &= 
    \frac{\tilde{\mu}_2}{2} v_C
    +\frac{\tilde{\mu}_1-\tilde{\mu}_3}{2} \frac{v_C^2+v_D^2}{v_C}
    -\tilde{m}^2
  \\
  M_{H_2^\pm}^2 &= 
    \frac{\tilde{\mu}_2}{2} v_C
    +\frac{\tilde{\mu}_1-\tilde{\mu}_3}{2} \frac{v_C^2+v_D^2}{v_C}
    +\tilde{m}^2
  \\
  M_{A^0}^2 &=
    \frac{v_D^2+v_C^2}{v_C} (b_7 v_C - \mu_3)
\end{align}
\end{subequations}

In the expressions in \eqref{eqn:CL-cust-decomp} and \eqref{eqn:CL-cust-masses} appear the rescaled parameters $\tilde\mu_i$ for the singly charged fields and the effective mass
\begin{equation}
  \tilde{m}^2 = 
    (\tilde{\mu}_1+\tilde{\mu}_2) \frac{v_C}{2} \frac{v_C^2+v_D^2}{v_C^2}
      \sqrt{1-\frac{(2\tilde{\mu}_1-\tilde{\mu}_2)(\tilde{\mu}_2+2\tilde{\mu}_3)}{(\tilde{\mu}_1+\tilde{\mu}_3)^2} \Bigl(\frac{v_C^2}{v_C^2+v_D^2}\Bigr)^2}
  \text{.}
\end{equation}
The $\tilde{\mu}_i$ are of the same size as the corresponding $\mu_i$ but receive some corrections of order of the EW scale.
The only exception is $\tilde{\mu}_2$ which gets an additional see-saw-like term $\sim v_D^2/v_C$.
However, this does not change the overall scale of the masses.
The exact definition of the $\tilde{\mu}_i$ can be found in \eqref{eqn:mass-sc-params}.

Like in the simultaneous limit all mass terms have a see-saw structure with the factor $v_D/v_C$.
But since $v_C$ does not need to be small compared to $v_D$ these non-SM like fields do not necessarily decouple in the custodial limit.
In that way the custodial limit offers the possibility of additional light scalar particles that may be detectable at the LHC.

In order to see that this is a viable option we still have to discuss the neutral CP-even states.
From the analytic expressions for the mass eigenvalues of the corresponding $3\times3$ mass matrix \eqref{eqn:mass-matrix-nc} it is not easy to infer that the eigenvalues are indeed positive.
Taking the limit $v_C \to 0$ we arrive at a special case of the simultaneous limit in which we derived the masses in \eqref{eqn:SL-masses}.
If we now continuously increase $v_C$ the obtained masses will also shift continuously and remain positive at least for moderate values of $v_C$.

To underline this behaviour we performed a numerical study of the 2HTM assuming the case of sufficient fine-tuning, i.e. only EWSB-scale dimensionful parameters.
For that we varied the model parameters in the ranges given in \autoref{tab:num-ints}.
The remaining mass parameters $m_D$, $m_T$ and $m_\Upsilon$ then were calculated using the minimisation conditions \eqref{eqn:ewsb-min-cond}.
For the VEVs we required that they give the right rho-parameter in \eqref{eqn:rho-parameter} and the right $W$- and $Z$-boson mass in \eqref{eqn:gauge-masses-ewsb}. 
\begin{table}[h]
\centering
  \begin{tabular}{lcc}
    \hline\\[-2ex]
      Parameter & Symbol & Range\\
    \hline\\[-2ex]
      dimensionless parameters & $a_i, b_i$ & $-5\ldots+5$\\
      doublet self-coupling & $l$ & $\phantom{-}0\ldots+5$\\
      trilinear couplings & $\mu_i$ & $-5\,\text{TeV}\ldots+5\,\text{TeV}$\\
    \hline\\[-2ex]
      doublet VEV $\langle\Phi_2\rangle$ & $v_D$ & $\phantom{-9}0\,\text{GeV}\ldots252\,\text{GeV}$\\
      triplet VEV $\langle\Sigma_3\rangle$ & $v_T$ & $-90\,\text{GeV}\ldots\phantom{9}90\,\text{GeV}$\\
      triplet VEV $\langle\Upsilon_3\rangle$ & $v_\Upsilon$ & $-90\,\text{GeV}\ldots\phantom{9}90\,\text{GeV}$\\[0.5ex]
    \hline
  \end{tabular}
  \caption{Parameter ranges for the simulation of the 2HTM}
  \label{tab:num-ints}
\end{table}

With the obtained parameter sets we calculated the scalar mass spectrum.
All combinations that led to negative mass values were rejected, since the EWSB vacuum is not stable in this case.                                                                                                                                                 
Furthermore we neglected all parameter combinations that lead to a too large mass splitting between $M_{A^0}$ and $M_{H^{\pm\pm}}$, since this indicates accidental fine-tuning in the numerical simulation.
In \autoref{tab:num-mass} we show the lower bounds on all scalar mass in dependence of the allowed SM-Higgs region.
\begin{table}[h]
\centering
  \begin{tabular}{l|cc|c}
    \hline
      \rule{0pt}{2.7ex} & \multicolumn{2}{c|}{custodial limit} & sim. limit\\
      Field & mass (LEP) & mass (LHC) & mass (LHC) \\
    \hline
      \rule{0pt}{2.7ex}\ $h^0$ & $>114.5\,\text{GeV}$ & $117-132\,\text{GeV}$ & $117-132\,\text{GeV}$\\
    \hline
      \rule{0pt}{2.7ex}\ $H^0_1$ & $>238\,\text{GeV}$ & $>234\,\text{GeV}$ & $>1738\,\text{GeV}$\\
      \ $H^0_2$ & $>424\,\text{GeV}$ & $>503\,\text{GeV}$ & $>1962\,\text{GeV}$\\
      \ $A^0$ & $>153\,\text{GeV}$ & $>221\,\text{GeV}$ & $>1738\,\text{GeV}$\\
    \hline
      \rule{0pt}{2.7ex}\ $H^{\pm\pm}$ & $>49\,\text{GeV}$ & $>129\,\text{GeV}$ & $>1784\,\text{GeV}$\\
      \ $H^\pm_1$ & $>136\,\text{GeV}$& $>136\,\text{GeV}$ & $>1761\,\text{GeV}$\\
      \ $H^\pm_2$ & $>377\,\text{GeV}$ & $>490\,\text{GeV}$ & $>1980\,\text{GeV}$\\[0.5ex]
    \hline
  \end{tabular}
  \caption{Lower mass bounds for the scalar fields of the 2HTM obtained by numerical simulation. The SM-Higgs-like scalar $h^0$ is restricted to the given range.}
  \label{tab:num-mass}
\end{table}

From \autoref{tab:num-mass} it is easy to infer that in the custodial limit much lighter masses for the investigated states are possible.
In all cases present experimental bounds are above the obtained values so they already restrict the parameter space of the model.
This is independent of whether the lightest SM Higgs like scalar is in range that is excluded by the LHC or not.

\section{Conclusions}
\label{sec:conclusion}
We investigated the complete scalar sector of an extended non-SUSY SU(5) model consisting of a \rep{5}, a \rep{15} and a \rep{24}.
Out of those multiplets at least the \rep{5} has to split into a light SM like Higgs doublet and a heavy scalar colour triplet to avoid excessive proton decay.
Starting from the resulting hierarchy we determined the masses of all scalar fields and examined the conditions that lead to light, i.e. EWSB scale, masses.
Gauge coupling unification favours light masses for $\Phi_2$, $\Sigma_3$, $\Upsilon_{LQ}$ and $\Upsilon_3$, but the mentioned doublet triplet splitting cannot explain a light leptoquark $\Upsilon_{LQ}$.
For this additional fine-tuning is needed.

Considering only the $\text{SU}(2)_L$-multiplets $\Phi_2$, $\Sigma_3$ and $\Upsilon_3$ we arrive at a generic two-Higgs-triplet model at the EWSB scale.
We demonstrated the decoupling of all additional scalar fields of this model from the SM in the simultaneous limit  $v_T, v_\Upsilon\to0$.
As an appealing feature of the 2HTM we investigated the possibility to maintain custodial symmetry through $v_T=v_\Upsilon$ (custodial limit).
In contrast to the simultaneously small $v_T$, $v_\Upsilon$ constraints from the EW gauge bosons and the rho-parameter do not push the masses of the new Higgs bosons above the known experimental limits (c.f. \autoref{tab:num-mass}, so that interesting scenarios for the LHC emerge.
Conversely the present mass bounds already put restrictions on the parameter space.
Since the SU(5) symmetry does not supply any parameter relations for the 2HTM this result holds beyond the studies GUT embedding and is valid for any two-Higgs-triplet model.

\section*{Acknowledgements}
I thank Ulrich Nierste for suggesting the topic, many fruitful discussions, and proofreading the manuscript.
I further appreciate many discussions with Martin Gorbahn and S\"oren Wiesenfeldt.
This work was partially funded by DFG through the Graduiertenkolleg \emph{Hochenergiephysik und Teilchenastrophysik} and by the German National Scholarship Foundation \emph{(Studienstiftung des deutschen Volkes)}.
\pagebreak

\appendix
\section{Details of the Potentials and Vevs} \label{app:potential}
The model under investigation comprises three scalar fields that transform under a \rep{5}, \rep{15}
and \rep{24} respectively. The usual notation of their decomposition into SM fields (c.f. 
\autoref{tab:higgses}) is given in \eqref{eqn:su5-2-sm-decomp}. 
\begin{subequations} \label{eqn:su5-2-sm-decomp}
\begin{gather}
\begin{aligned}
  \Phi &= \begin{pmatrix} \Phi_3\\ \Phi_2 \end{pmatrix} &\qquad
  \Upsilon &=
    \begin{pmatrix} \Upsilon_6 & \Upsilon_{LQ}\\ \Upsilon_{LQ} & \Upsilon_3\end{pmatrix}
\end{aligned}\\
  \Sigma = 
    \Biggl(\begin{matrix} 
      \Sigma_8 & \Sigma_{NGB}\\ \overline{\Sigma}_{NGB} & \Sigma_3
    \end{matrix}\Biggr)
    + \frac{\Sigma_1}{2} 
      \Biggl(\begin{smallmatrix} 2\\&2\\&&2\\&&&-3\\&&&&-3\end{smallmatrix}\Biggr)
\end{gather}
\end{subequations}
Using the symmetry of $\Upsilon$ and the hermiticity of $\Sigma$ the most general renormalizable
potential is derived in \eqref{eqn:potential}. Note that the first two lines cover the minimal
SU(5) model of Georgi and Glashow \cite{Georgi:1974sy} following \cite{Buras:1977yy} as a
convention for the parameters.
\begin{equation} \label{eqn:potential}
\begin{aligned}
\mathcal{V} &= \frac{a_1}{4} \bigl(\tr\Sigma^2\bigr)^2 
    + \frac{b_1}{2} \tr\Sigma^4 
    + c_1 \tr\Sigma^3
    - m_1^2 \tr\Sigma^2
    + \frac{l}{4} \bigl(\Phi^\dag \Phi\bigr)^2
    - m_2^2\,\Phi^\dag \Phi
\\
    &\quad
    + a_2\,\Phi^\dag\Phi \tr\Sigma^2 
    + b_2\,\Phi^\dag\Sigma^2\,\Phi 
    + c_2\,\Phi^\dag\Sigma\,\Phi
\\
    &\quad
    + \frac{a_3}{4} \bigl(\tr \Upsilon^\dag\Upsilon\bigr)^2
    + \frac{b_3}{2} \tr\bigl(\Upsilon^\dag\Upsilon\bigr)^2
    - m_3^2 \tr \Upsilon^\dag \Upsilon
\\
    &\quad + \frac{c_4}{2} \bigl(\Phi \Upsilon^\dag\Phi + \hc\bigr)
    + c_5 \tr \Upsilon^\dag \Sigma \Upsilon
    + a_4\,\Phi^\dag \Phi \tr \Upsilon^\dag \Upsilon
    + a_5 \tr \Upsilon^\dag \Upsilon \tr \Sigma^2 
\\
    &\quad + b_4\,\Phi^\dag \Upsilon \Upsilon^\dag \Phi
    + b_5 \tr \Upsilon^\dag \Sigma^2 \Upsilon
    + b_6 \tr \Upsilon^\dag \Sigma \Upsilon \Sigma^T
    + b_7 \bigl(\Phi \Upsilon^\dag \Sigma \Phi + \hc\bigr)
\end{aligned}
\end{equation}
In the notation of \eqref{eqn:su5-2-sm-decomp} the VEVs are chosen to take the form
\begin{subequations}
\begin{gather}
\langle\Sigma_1\rangle = v_S \diag\bigl(1,1,1,-\tfrac{3}{2},-\tfrac{3}{2}\bigr)\text{,}\\
\begin{aligned}
  \langle\Phi_2\rangle &= 
    \frac{v_D}{\sqrt{2}} \begin{pmatrix} 0\\ 1\end{pmatrix}\text{,} &
  \langle\Sigma_3\rangle &=
    \frac{v_T}{2} \begin{pmatrix} 1 & 0\\ 0 & -1\end{pmatrix}\text{,} &
  \langle\Upsilon_3\rangle &= 
    v_\Upsilon \begin{pmatrix} 0 & 0\\ 0 & 1\end{pmatrix}\text{.}
\end{aligned} \label{eqn:vev-ewsb}
\end{gather}
\end{subequations}

As was shown in \autoref{sec:spectrum} the SU(5) breaking due to $\langle\Sigma_1\rangle$ leads to heavy masses for all scalars except $\Phi_2$, $\Sigma_3$ and $\Upsilon_3$ due to constraints from proton decay. 
Therefore it is possible to construct an effective model comprising only these fields.
Its potential is given in \eqref{eqn:potential-2htm}, where the notation of \eqref{eqn:vev-ewsb} is used omitting the component indices.
\begin{equation} \label{eqn:potential-2htm}
\begin{aligned}
\mathcal{V}_\text{2HTM} &= 
  \frac{m_D^2}{4} \Phi^\dag \Phi
  + \frac{m_T^2}{4} \tr \Sigma^2
  + \frac{m_\Upsilon}{4} \tr\Upsilon^\dag \Upsilon\\
&\quad
  + \mu_1 \Phi^\dag \Sigma \Phi
  + \mu_2 \tr \Upsilon^\dag \Sigma \Upsilon
  + \frac{\mu_3}{2} \bigl(\Phi \Upsilon^\dag \Phi +\hc\bigr)\\
&\quad
  + \frac{l}{4} \bigl(\Phi^\dag \Phi\bigr)^2
  + \frac{a_1+b_1}{4} \bigl(\tr \Sigma^2\bigr)^2
  + \frac{a_3}{4} \bigl(\tr\Upsilon^\dag \Upsilon\bigr)^2
  + \frac{b_3}{2} \tr\bigl(\Upsilon^\dag \Upsilon\bigr)^2\\
&\quad
  + b_4 \Phi^\dag \Upsilon \Upsilon^\dag \Phi
  + b_6 \tr \Upsilon^\dag \Sigma \Upsilon \Sigma^T
  + b_7 \bigl(\Phi \Upsilon^\dag \Sigma \Phi + \hc\bigr)\\
&\quad
  + \frac{2 a_2+b_2}{2} \Phi^\dag \Phi \tr \Sigma^2
  + a_4 \Phi^\dag \Phi \tr \Upsilon^\dag \Upsilon
  + \frac{2a_5+b_5}{2} \tr\Sigma^2 \tr\Upsilon^\dag\Upsilon
\end{aligned}
\end{equation}
In \eqref{eqn:eff-fields} we set a notation for the field components utilizing their electric charges.
\begin{subequations} \label{eqn:eff-fields}
\begin{gather} 
  \begin{aligned}
  \Phi_2 &= 
    \begin{pmatrix}
      \Phi_2^+\\
      \Phi_2^0
    \end{pmatrix} &
  \Phi_2^\dag &= 
    \begin{pmatrix}
      \Phi_2^0{}^*\\
      \Phi_-
    \end{pmatrix} &
  \Sigma_3 &= \frac{1}{2}
    \begin{pmatrix}
      \Sigma_3^0 & \Sigma_3^+\\
      \Sigma_3^- & -\Sigma_3^0
    \end{pmatrix}
  \end{aligned}\\
  \begin{aligned}
  \Upsilon_3 &=
    \begin{pmatrix}
      \Upsilon_3^{++} & \Upsilon_3^+/\sqrt{2}\\
      \Upsilon_3^+/\sqrt{2} & \Upsilon_3^0
    \end{pmatrix} &
  \Upsilon_3^\dag &=
    \begin{pmatrix}
      \Upsilon_3^{--} & \Upsilon_3^-/\sqrt{2}\\
      \Upsilon_3^-/\sqrt{2} & \Upsilon_3^0{}^*
    \end{pmatrix}
  \end{aligned}
\end{gather}
\end{subequations}

The components of the same charges mix into the mass eigenstates which can be obtained by diagonalising the respective matrices given in \eqref{eqn:mass-matrices}.
\begin{subequations} \label{eqn:mass-matrices}
\begin{align}
  \mathcal{M}_3^\pm &= 
    \begin{pmatrix}
      \frac{v_D^2}{v_T} \tilde{\mu}_1 + \frac{1}{2} \frac{v_\Upsilon^2}{v_T} \tilde{\mu}_2 &
      \mp\im v_D \tilde{\mu}_1 &
      \mp\frac{\im}{2} v_\Upsilon \tilde{\mu}_2\\
      \pm\im v_D \tilde{\mu}_1 & 
      v_T \tilde{\mu}_1 - v_\Upsilon \tilde{\mu}_3 &
      v_D \tilde{\mu}_3\\
      \pm\frac{\im}{2} v_\Upsilon \tilde{\mu}_2 &
      v_D \tilde{\mu}_3 &
      \frac{1}{2} \tilde{\mu}_2 v_T - \frac{v_D^2}{v_\Upsilon} \tilde{\mu}_3
    \end{pmatrix}\\
\label{eqn:mass-matrix-nc}
  \mathcal{M}_3^0 &= 
    \begin{pmatrix}
      0 &
      -v_D \bar{\mu}_1 &
      \sqrt{2} v_D \bar{\mu}_3\\
      -v_D \bar{\mu}_1 &
      \frac{v_D^2}{v_T} \bar{\mu}_1 + \frac{v_\Upsilon^2}{v_T} \frac{\bar{\mu}_2}{2} &
      -\frac{1}{\sqrt{2}} v_\Upsilon \bar{\mu}_2\\
      \sqrt{2} v_D \bar{\mu}_3 &
      -\frac{1}{\sqrt{2}} v_\Upsilon \bar{\mu}_2 &
      - \frac{v_D^2}{v_\Upsilon} \bar{\mu}_3
    \end{pmatrix}\nonumber\\
  &\quad +
    \begin{pmatrix}
      2 v_D^2 l &
      0 & 
      0\\
      0 &
      \parbox{11em}{$
      v_D^2(2 a_2+b_2)
      + v_T^2 \frac{a_1+b_1}{2}\\
      + v_\Upsilon^2 \frac{2a_5+b_5+b_6}{2}
      - v_D^2\frac{v_\Upsilon}{v_T} b_7$} &
      0\\
      0 & 
      0 & 
      \parbox{6em}{$
      2 v_D^2 (a_4+b_4)\\
      + v_\Upsilon^2 (a_3+2b_3)$}
    \end{pmatrix}
    +\mathcal{O}(\epsilon^2v_S^2)\\
  \mathcal{M}_2^0 &= 
    \begin{pmatrix}
      -2 \frac{v_\Upsilon}{v_D}  & \sqrt{2}\\
      \sqrt{2} & - \frac{v_D}{v_\Upsilon}
    \end{pmatrix}(\mu_3 - v_T b_7)\,v_D
\end{align}
\end{subequations}
In the matrices $\mathcal{M}_3^\pm$ and $\mathcal{M}_3^0$ of the singly charged and neutral CP-even fields effective parameters $\tilde{\mu}_i$ and $\bar{\mu}_i$ are used.
They are related to the couplings via \eqref{eqn:mass-sc-params}.
\begin{subequations}  \label{eqn:mass-sc-params}
\begin{align}
  \tilde{\mu}_1 &= \mu_1 + b_7 v_\Upsilon &
  \bar{\mu}_1 &= \mu_1 - (2 a_2 + b_2) v_T + 2 b_7 v_\Upsilon\\
  \tilde{\mu}_2 &= \mu_2 + 2 b_7 \frac{v_D^2}{v_\Upsilon} - b_6 v_T &
  \bar{\mu}_2 &= \mu_2 + 2 b_7 \tfrac{v_D^2}{v_\Upsilon}\\
  \tilde{\mu}_3 &= \mu_3 + b_4 v_\Upsilon &
  \bar{\mu}_3 &= \mu_3 - b_7 v_T + 2 (a_4 + b_4) v_\Upsilon
\end{align}
\end{subequations}
The expansion parameter $\epsilon$ in \eqref{eqn:mass-matrices} describes the hierarchy between GUT and EW scale as indicated in section~\ref{sec:model}.
The eigenvalues and states of these matrices are discussed in section~\ref{sec:sim-limit} and section~\ref{sec:cust-limit}.
\pagebreak

\bibliography{paper1}
\bibliographystyle{JHEP}

\end{document}